\documentclass[pra,amsfonts,twoside,amssymb,superscriptaddress,twocolumn]{revtex4}

\usepackage{graphicx, color, graphpap}
\usepackage{amsmath}
\usepackage{enumitem}
\usepackage{amssymb}
\usepackage{amsthm}
\usepackage{pstricks}
\usepackage{float}
\usepackage{multirow}
\usepackage{refcount}

\newcommand{\ket}[1]{\vert #1\rangle}    
 
\renewcommand{\>}{\rangle}
\newcommand{\ot}{\otimes}
\newtheorem{Def}{Definition}[]
\newtheorem{Thm}[Def]{Theorem}
\newtheorem{Lem}[Def]{Lemma} 
\newtheorem{Ex}[Def]{Example}
\newtheorem{Prop}[Def]{Proposition}
\newcommand{\bdf}{\begin{Def}}
\newcommand{\edf}{\end{Def}}
\newcommand{\bex}{\begin{Ex}}
\newcommand{\eex}{\end{Ex}}
\newcommand{\bthm}{\begin{Thm}}
\newcommand{\ethm}{\end{Thm}}
\newcommand{\blm}{\begin{Lem}}
\newcommand{\elm}{\end{Lem}}
\newcommand{\bprop}{\begin{Prop}}
\newcommand{\eprop}{\end{Prop}}

\def\ket #1{\vert #1\rangle}

\newcommand*{\cD}{\mathcal{D}}

\newcommand*{\cH}{\mathcal{H}}

\newcommand*{\cJ}{\mathcal{J}}

\newcommand*{\cL}{\mathcal{L}}

\newcommand*{\cO}{\mathcal{O}}

\newcommand*{\tr}{\mathrm{tr}}

\definecolor{myred}{rgb}{1,0,0}

\definecolor{myblue}{rgb}{0,0,0.8}

\definecolor{myyellow}{rgb}{0.9,0.8,0}

\definecolor{mygreen}{rgb}{0,0.6,0}

\definecolor{myorange}{rgb}{0.6,0.6,0}

\definecolor{mycerul}{rgb}{0,0.6,1}

\begin{document}

\title{Information-theoretical formulation of anyonic entanglement}

\author{Kohtaro Kato}
\affiliation{Department of Physics, Graduate School of Science, The University of Tokyo, Tokyo, Japan}

\author{Fabian Furrer}
\affiliation{Department of Physics, Graduate School of Science, The University of Tokyo, Tokyo, Japan}

\author{Mio Murao}
\affiliation{Department of Physics, Graduate School of Science,
The University of Tokyo, Tokyo, Japan}
\affiliation{Institute for Nano Quantum Information Electronics,
The University of Tokyo, Tokyo, Japan}

\date{\today}

\begin{abstract}

Anyonic systems are modeled by topologically protected Hilbert spaces which obey complex superselection rules restricting possible operations. These Hilbert spaces cannot be  decomposed into tensor products of spatially localized subsystems, whereas the tensor product structure is a foundation of the standard entanglement theory. 
We formulate bipartite entanglement theory for pure anyonic states and analyze its properties as a non-local resource for quantum information processing. We introduce a new entanglement measure, asymptotic entanglement entropy (AEE), and show that it characterizes distillable entanglement and entanglement cost similarly to entanglement entropy in conventional systems. AEE depends not only on the Schmidt coefficients but also on the quantum dimensions of the anyons shared by the local subsystems. Moreover, it turns out that AEE coincides with the entanglement gain by anyonic excitations in certain topologically ordered phases.
\end{abstract}
\maketitle

\section{Introduction}
The discovery of topologically ordered phases as in the fractional quantum hall effect~\cite{PhysRevLett.48.1559, PhysRevLett.50.1395, PhysRevLett.53.722} has revealed a new kind of quantum phase not described by the conventional symmetry breaking picture.  In topologically ordered phases, energy eigenstates have distinct properties, namely, the ground states have topological degeneracy and the statistics of quasiparticle excitations are not necessarily fermonic or bosonic, but {\it anyonic}.  Much progress has recently been made in understanding quantum properties of topologically ordered phases~(see e.g., Ref.~\cite{Wenrev2013} and references therein). In particular, the analysis by means of entanglement -- a standard procedure in quantum information theory~\cite{RevModPhys.81.865} -- revealed new properties distinguishing topologically ordered phases from conventional phases. 
In Refs.~\cite{PhysRevLett.96.110404,PhysRevLett.96.110405}, it has been shown that the ground states of topologically ordered phases exhibit specific patterns of long-range entanglement that can be characterized by the so-called topological entanglement entropy (TEE). Recently, TEE has also been extended to states including anyonic excitations in topologically ordered phases~\cite{2008JHEP...05..016D,PhysRevLett.111.220402,2014arXiv1403.0702H,2008AnPhy.323.1729H}.

However, entanglement is not only useful to classify states, but also has operational relevance in quantum information theory. It is for instance the resource that enables quantum teleportation~\cite{PhysRevLett.70.1895} or measurement-based quantum computation~\cite{PhysRevLett.86.5188}. This aspect of entanglement is particularly interesting since anyonic excitations are robust against local noise, making them promising candidates for fault-tolerant quantum information processing by encoding quantum information in the topologically protected Hilbert space (also called fusion space)~\cite{Kitaev2003a,Freedman2002,RevModPhys.80.1083}. This gives rise to the question whether TEE of anyonic excitations of topologically ordered phases allows an operational interpretation for fault-tolerant quantum information processing. In order to address this question, we consider a resource theoretical analysis of bipartite pure entanglement of anyonic states in the topologically protected Hilbert space.

The standard entanglement theory cannot be directly applied to anyonic systems since the topologically protected Hilbert space describing anyons cannot be decomposed into tensor products of local subsystems. This property can be understood as the existence of superselection rules restricting  physically possible operations.  Entanglement properties under superselection rules induced by group symmetries or superselection rules of fermions have been recently studied in Refs.~\cite{PhysRevLett.91.097903, PhysRevLett.92.087904, PhysRevA.87.022338}, but their techniques cannot be applied to anyons.

In this paper, we formulate an entanglement theory for anyonic systems in pure bipartite settings from an information theoretical viewpoint, and find a qualitatively different behavior compared to ordinary systems with tensor product structure. In the standard entanglement theory, the uniqueness theorem \cite{donald2002uniqueness} states that many operationally defined entanglement measures  coincide to the entanglement entropy (EE) in an asymptotic situation of infinite identical copies due to the additivity of EE.  Distillable entanglement and entanglement cost are examples of such operational measures. 

However, we find that for anyonic systems the corresponding EE is not additive as required for an operational mesaure in the asymptotic limit, but super-additive. We then show that the asymptotic rate of EE which we call asymptotic entanglement entropy (AEE) is given by EE plus a non-negative term that depends on the quantum dimension of the shared anyon charge. 
Moreover, we prove that AEE corresponds to the distillable entanglement and the entanglement cost in anyonic systems establishing an operational interpretation of AEE.

Interestingly, we find that the operationally meaningful AEE coincides with the contribution of anyonic excitations in topologically ordered phases to TEE~\cite{2008JHEP...05..016D,PhysRevLett.111.220402, 2014arXiv1403.0702H}, which has been derived by Hikami~\cite{2008AnPhy.323.1729H} for topologically ordered phases that can be described by $SU(2)_k$ Chern-Simons theories~\cite{1974CS,Wittentqft}. This shows that all the topological entanglement that is not contained in the ground state can be distilled by only braiding operations and can therefore be exploited for fault-tolerant information processing. 

The paper is organized as follows. In Sec.~\ref{sec.2},  we introduce a general model to describe anyonic systems. We formally define entangled states in anyonic systems in Sec.~\ref{sec.3} and introduce the AEE in Sec.~\ref{sec.4}. In Sec.~\ref{sec.5}, we derive basic properties of the AEE, and in Sec. VI we eventually present the equivalence between the AEE and entanglement of distillation and entanglement cost. Finally we conclude our findings in Sec. VII. Appendices contain details of the proof of Theorem~\ref{thm:distillcost} and a possible way to extend our results to more general anyonic systems which have non-vacuum total charge.

\section{Anyon Models} \label{sec.2}
A general anyon model is determined by its possible charges (i.e.,~anyon types), fusion rules and  braiding statistics. In the following we assume that the possible charges are labeled by a finite set $\cL=\{1,a,b,c,...\}$ where $1$ denotes the unique vacuum. The number of ways for $a$ and $b$ fusing to $c$ is given by $N_{ab}^c\in\mathbb{N}$ and we simply write $a  \times b = \sum_c N_{ab}^c \,  c$ to indicate the possible fusion channels of $a$ and $b$. For any $a\in \cL$, the vacuum satisfies $1\times a = a $  and there exists a unique anti-charge $\bar a$ such that $N_{a{\bar a}}^1=1$.  A charge $a$ is said to be abelian if the fusion ways for any $b$ is unique $\sum_c N_{ab}^c = 1$, and otherwise non-abelian. Moreover, we call a charge $a$ primitive if the corresponding matrix $(N_a)_{ij}=N_{ai}^j$ is primitive, i.e., for large $n$, $(N_{a}^n)_{ij}>0$ for all $i$ and $j$. One of the most studied primitive anyon due to its simplicity is the Fibonacci anyon $\tau$ in the Fibonacci anyon model $\{1,\tau\}$ which allows for universal quantum computation~\cite{Freedman2002}
and has quantum dimension $d_\tau=\frac{1+\sqrt{5}}{2}$. 
We note further that the so-called Ising anyon in the Ising anyon model $\{1,\sigma,\psi\}$ is non-abelian but not primitive. Moreover, the Ising model does not enable universal quantum computation~\cite{Freedman2002}.  

\begin{figure}
\includegraphics[width=\hsize]{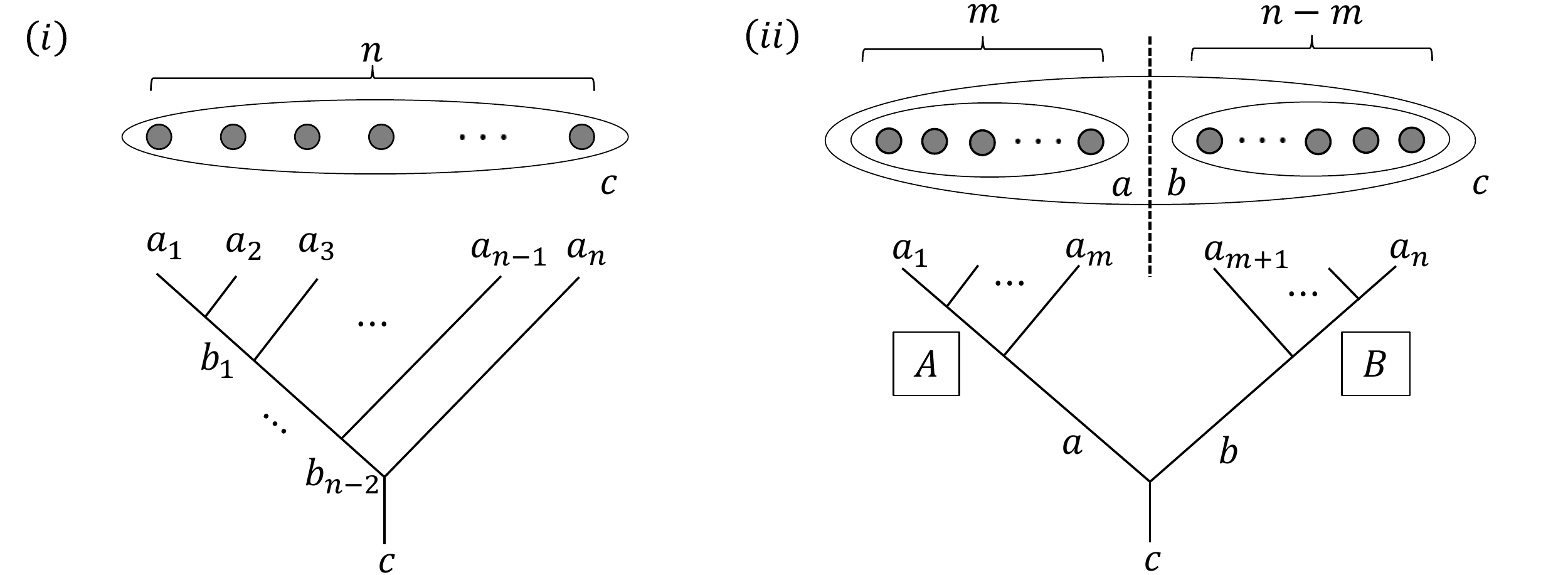}
\caption{\label{basisch}Tree representation for two different bases of the fusion space of the anyon chain $\{a_1,a_2,...,a_n\}$. 
The left hand side corresponds to a decomposition~\eqref{eq:fspaceChain}, where $b_1,...,b_{N-2}$ parametrize the different fusion outcomes. The right hand side shows a basis corresponding to the decomposition in~\eqref{eq:Decomp2}, where $a$ and $b$ denote the total charge of each subsystem. }
\end{figure}

The Hilbert spaces associated to an anyon model are called fusion spaces and spanned by the possible fusion ways. In particular, the fusion space of two charges $a$ and $b$ fusing to $c$ is given by
\begin{equation}\label{def:fusionspace}
V_{ab}^c={\rm span}\{\vert a b : c,\mu \rangle \, | \, \mu=1,2,...,N_{ab}^c\} \, .  
\end{equation}
Note that the order of the anyons and its arrangement on the two dimensional manifold is crucial because of the braiding statistics (c.f. Ref.~\cite{PhysRevB.89.035105}). In this paper, we use the convention that any anyon chain $\{a_1,a_2,...,a_n\}$ with total charge $c$ is assumed to be consecutively aligned from left to right on a two dimensional disc. The corresponding fusion space of such a chain can be constructed from two-anyon fusion spaces~(\ref{def:fusionspace}) as
\begin{equation}\label{eq:fspaceChain}
V_{a_1...a_n}^c=\bigoplus_{b_1... b_{n-2}} V_{a_1a_2}^{b_1}\otimes V_{b_1a_3}^{b_2} \otimes \cdots \otimes V_{b_{n-2}a_n}^c  \, .
\end{equation} 
The dimension of $V^c_{a_1...a_n}$ is given by
\begin{equation}\label{dimfu}
\dim V^c_{a_1...a_n}=\sum_{b_1, b_2,..., b_{n-2}} N_{a_1a_2}^{b_1}N_{b_1a_3}^{b_2}...N^{c}_{b_{n-2}a_n}.
\end{equation}
It is convenient to illustrate the decomposition of the fusion space by tree diagrams (Fig.~\ref{basisch}$(i)$).
The scaling behavior of the dimension of  the fusion space $V_{a^n}^c\equiv V_{a...a}^c$ of  $n$ $a$-anyons $\{a,a,...,a\}$ is determined by the {\it quantum dimension} $d_a$ which satisfies  $d_ad_b=\sum_{c}N_{ab}^cd_c$ for all $a,b,c\in \cL$. 
For example, the quantum dimension of an abelian anyon $a$ is $d_a=1$, and $\dim V_{a^n}^c=1$ if $a^n$ can fuse to $c$ and $0$ else. 
If $a$ is a primitive charge, then for any $b\in \cL$ it holds that~\cite{Verlinde1988360,Preskill2004}
\begin{equation}\label{qudimn}
\dim V_{a^n}^b=\frac{d_a^nd_b}{{\cal D}^2}\left(1+{\cal O}(c_b^n)\right) \, ,
\end{equation} 
where $\vert c_b \vert < 1$ and $\cD$ denotes the total quantum dimension given by $\cD=\sqrt{\sum_ad_a^2}$. If $a$ is not primitive, the above formula does not hold since the fusion channel $a^n\to b$ does not always exist for any $b$ and large $n$.  
However, it still holds that $\dim V_{a^n}^b\propto d_a^n$ for sufficiently large $n$ if the dimension is not zero. 
More generally if at least $a$ is primitive, it holds that 
\begin{equation}\label{qudimnprimitive}
\dim V_{a^{n_a}b^{n_b}... 1^{n_1}}^x=\frac{d_a^{n_a}d_b^{n_b}...d_1^{n_1}d_x}{{\cal D}^2}\left(1+{\cal O}(c_x^{n_a})\right) \,,
\end{equation}
where $|c_x|<1$. The proof is similar to the one of Eq.~\eqref{qudimn}. We emphasize that commutativity of the fusion matrix is essential since the proof of these relations requires the Perron-Frobenius theorem. 

The quantum dimensions satisfy the following property which we will use later. 
\blm\label{dimdim}
For all $n\in \mathbb{N}$ holds that
\begin{equation}
d_{a_1}...d_{a_n}=\sum_b\dim V_{a_1...a_n}^bd_b
\end{equation}
\elm
\begin{proof}
 Using the equation $d_ad_b=\sum_cN_{ab}^cd_c$ recursively, we can compute 
 \begin{align}
 d_{a_1}d_{a_2}...d_{a_n}&=\sum_{b_1}N_{a_1a_2}^{b_1}d_{b_1}d_{a_3}...d_{a_n}\\
&=\sum_{b_1,b_2}N_{a_1a_2}^{b_1}N_{b_1a_3}^{b_2}d_{b_2}d_{a_4}...d_{a_n}\\
 &=\sum_{b_1,...,b_{n-2},c}N_{a_1a_2}^{b_1}N_{b_1a_3}^{b_2}...N_{b_{n-2}a_n}^cd_c \, .
 \end{align}
Hence, inserting relation \eqref{dimfu} completes the proof. 
\end{proof} 

\begin{figure}[t]
\includegraphics[width=1.0\hsize]{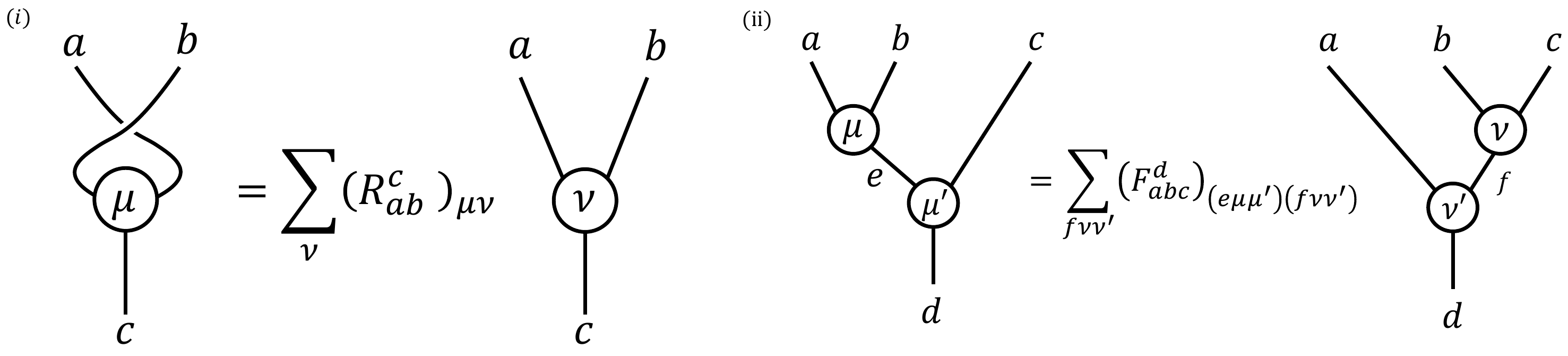}
\caption{\label{RF} (i) A graphical tree representation of the R-matrix $R_{ab}^c$ transforming $V_{ba}^c$ into $V_{ab}^c$. The tree on the left hand side of the equality represents the vector $\ket{ba;c,\mu}$ and the sum over $\nu$ on the right hand side goes over all trees representing vectors $\ket{ab;c,\nu}$. 
(ii) A graphical tree representation of the F-matrix $F_{abc}^d$. The tree on the left hand side of the equality represents the vector $\ket{ab;e,\mu}\otimes\ket{ec;d,\mu '}$. This vector can be expanded as a superposition of vectors of the form $\ket{af;d,\nu'}\otimes\ket{bc;f,\nu}$, where the coefficients of the superposition are determined by the F-matrix $F_{abc}^d$. 
}
\end{figure}

Interchanging neighboring anyons acts as a unitary transformation $R_{ab}^c: V_{ab}^c \rightarrow V_{ba}^c$ and is called the R-matrix. 
Due to the associativity of the fusion rule $ (a\times b)\times c = a\times (b\times c)$, there exist also isomorphisms $F_{abc}^d$ relating the fusion spaces $\bigoplus_x V_{ab}^{x} \otimes V_{xc}^d$ and $\bigoplus_x V_{ax}^{d} \otimes V_{bc}^x$ referred to as the F-matrix. An arbitrary braiding operation is then fully described via R- and F-matrices (Fig.~\ref{RF}).

\section{Entangled States in Anyonic Systems}\label{sec.3}
In the following, we investigate bipartite entanglement of a pure state of an anyon chain $\{a_1,a_2,... ,a_n\}$ on a disc.
As elaborated in Ref.~\cite{PhysRevB.89.035105}, entanglement has to be defined according to a splitting of the disc into submanifolds. In this paper, we focus on the situation where the locations of the parties are spatially well separated and divide the disc into two halves $A$ and $B$ that contain the anyons $\{a_1,...,a_m\}$ and $\{a_{m+1},...,a_n\}$, respectively (Fig.\ref{basisch}(ii)). This implies that neither $A$ nor $B$ is allowed to move anyons along a path enclosing anyons belonging to the other party.  

We denote by $\cH_A^x$ and $\cH_{B}^x$ the fusion space of the anyon chain $A$ and $B$ with total charge $x$.  
Then, the joint fusion space $\cH_{AB}^c := V^c_{a_1...a_n}$ with total charge $c$ can be decomposed as~\cite{kitaev2004superselection}
\begin{equation}\label{eq:Decomp2}
\cH_{AB}^c=\bigoplus_{a,b\in{\cal L}}\cH_A^a\ot\cH_B^b\ot V_{ab}^c\,,
\end{equation}
with a tree diagram as depicted in Fig.~\ref{basisch}(ii). Hence, the joint Hilbert space can generally not be written as the tensor product of the individual subsystems $A$ and $B$.

We are now interested in the state restricted to the local subsystem $A$. 
A superselection rule restricts the possible operations on the subsystem $A$ such that the total charge of $A$ is conserved~\cite{kitaev2004superselection}.
Hence, the Hilbert space corresponding to $A$ is composed of superselection (SS) sectors each determined by the local total charge $a$, and the reduced state can be assumed to have block structure with respect to $a$. 
This motivates to define a partial trace on anyonic systems in the following way. We first embed $\cH_{AB}^c$ in the larger (non-physical) Hilbert space ${\cH}_{AB}={\cH_A}\ot {\cH}_{\bar A}$,
where ${\cH}_A=\bigoplus_a\cH_A^a$ and ${\cH}_{\bar A}=\bigoplus_{ab}\cH_B^b\ot V_{ab}^c$.  This canonical embedding and its natural extension to operators will be denoted by $\cJ$. 
We then define the partial trace on the anyonic system mapping states on $\cH_{AB}^c$ to states on $\cH_{A}$ by $\tau_{\bar A}=\tr_{\bar A}\circ\cJ$, where $\tr_{\bar A}$ denotes the usual partial trace over the system ${\cH}_{\bar A}$. By construction it is clear that the corresponding reduced states $\rho_A$ have  block structure $\rho_A = \bigoplus_a  \rho^a_A$, which will be used below. 
Note that if the total charge $c$ is the vacuum $1$, we have  ${\cH}_{\bar A}={\cH}_B$.

We define operations on an anyon chain as any combination of ($i$) adding ancillary anyons with total charge $1$, ($ii$) tracing out a part of the anyon chain, ($iii$) applying unitary transformations by braiding neighboring anyons and ($iv$) projective measurements which respect superselection rules. Local operations on $A$ and $B$ are given by operations only including anyons from $A$ and $B$, respectively. 
This means that only braiding between anyons inside the respective submanifold is allowed. 
In analogy to tensor product systems, we define a separable state as a state which can be created from $|\psi_A^1\>|\psi_B^1\>$ by using local operation and classical communication (LOCC), where $|\psi^1_{A}\>(|\psi^1_B\>)$ is a pure state on $\cH_{A}^1(\cH_B^1)$. Otherwise, a state is called entangled. Since the total charge of two systems is only uniquely determined if one of the total charges of the subsystems is abelian, this definition can be formulated as follows.
\bdf 
A bipartite pure state $|\psi_{AB}\>\in \cH_{AB}^c=\bigoplus_{ab}\cH_A^{a}\ot \cH_B^{b}\ot V_{ab}^c$  in an anyonic system is separable if there exist states $|\psi^a_A\>\in \cH_A^a$ and $|\psi^b_B\>\in \cH_B^b$ such that $|\psi_{AB}\>=|\psi^a_A\>|\psi^b_B\>$
 and at least one of the anyons $a$ or $b$ is abelian (thus, $V_{ab}^c$ is one-dimensional). A mixed state on $\cH_{AB}^c$ is called separable if it can be written as a convex combination of separable pure states. Moreover, a state which is not separable is called entangled.
\edf

\section{Asymptotic Entanglement Entropy}\label{sec.4}
Given a qudit system $\cH_A\ot \cH_B={\mathbb C}^d\times {\mathbb C}^d$, the entanglement entropy $E_A$ between $A$ and $B$ of a pure state $\rho $ is defined as the von Neumann entropy of the reduced state $\rho_A=\tr_B\rho$, that is, $E_A(\rho)  = -\tr\rho_A\log_2 \rho_A$ (we will use 2 as the base of $\log$). 
In anyonic systems, we define a generalization of the qudit entanglement entropy for a pure state $\rho=|\psi\>\langle \psi|$ on $\cH^c_{AB}$ by 
 \begin{equation}
 E_A^1(\rho)\equiv -\tr \rho_A\log\rho_A \, ,
 \end{equation}
 where $\rho_A=\tau_{\bar A} (|\psi\>\langle \psi|)$. 
Note that $E_A^1$ is equivalent to the entanglement entropy of anyonic systems defined in Ref.~\cite{PhysRevB.89.035105}. (Therein,  the quantum trace is used which corresponds to the trace in the fusion space with appropriate normalization for each superselection sector~\cite{2008AnPhy.323.2709B}.)

In the following we use the notation $E_A^1(|\psi\>)$ interchangeably with $E_A^1(\rho)$. 
In the case of $c=1$, any pure state can be written as $|\psi\>=\sum_a\sqrt{p_a}|\psi_a\>$, where $|\psi_a\>\in \cH_A^a\ot \cH_B^{\bar a}$, and it follows that
\begin{align}
E_A^1(|\psi\>)=H(\{p_a\})+\sum_a p_aE_A(|\psi_a\>) \, ,
\end{align}
where $H(\{p\})$ denotes the Shannon entropy of a probability distribution $p$. 
Note that it is crucial here that $\rho_A=\bigoplus_a\rho_A^a$ has block structure such that the distribution $\{p_a\}$ over the sectors are treated as purely classical degrees of freedom.

According to quantum information theory, the von Neumann entropy attains its operational significance in the asymptotic limit of an infinite number of independent and identical copies of the state. 
In an anyon model, independent and identical copies correspond to independent preparation of identical states in the same anyonic system. As discussed before, however, the fusion space corresponding to $N$ copies  cannot be written as the $N$-fold tensor product of the single copy fusion spaces. Moreover, in order to unambiguously define the $N$-copy of an anyonic state $\rho$, we have to assume that its total charge is vacuum (or abelian), otherwise no unique fusion channel exists (see Appendix~\ref{sec:nonvac} for a discussion of possible extensions). Since the total charge of the $N$-copy state is also  vacuum, the relevant fusion space is given by $\cH^1_{A^NB^N} =\bigoplus_{ a}\cH_{A^N}^a\ot\cH_{B^N}^{\bar a}$, where 
 we omit the trivial spaces $V_{a{\bar a}}^1$.
Here, the Hilbert space $\cH_{A^N}^a$, and similar $\cH_{B^N}^b$, is defined as
\begin{equation}
\cH_{A^N}^a\equiv\bigoplus_{a_1,...,a_N}\cH_{A_1}^{a_1}\ot\cdots\ot\cH_{A_N}^{a_N}\ot V_{a_1...a_N}^a ,
\end{equation}
where $A_i$ denotes a subsystem in $A$ corresponding to the $i$th copy. 
 We then define the $N$-copy state of $\rho$ by $\rho^N=\iota(\rho^{\otimes N})$, where $\iota$ denotes the corresponding embedding of $\cH_{AB}^{\otimes N}$ into $\cH^1_{A^NB^N}$ (and similarly its extension to the state space).

The fact that $\rho^N$ is not equivalent to the $N$-fold tensor product of $\rho$, and thus, neither is $\rho_A^{\otimes N}$ of $\rho_A =\tau_B(\rho)$, implies that $E^1_A(\rho^N)$ is generally not equal to $NE^1(\rho)$. This motivates to define the following asymptotic version of $E_A^1$.

\bdf
The asymptotic entanglement entropy (AEE) of an anyonic pure bipartite state $\rho$ is defined as
\begin{equation}
E^\infty_A(\rho) = \lim_{N\to\infty}\frac{E_A^1(\rho^N)}{N} \, .
\end{equation}
\edf

\begin{figure}[tb]
\begin{center}
\includegraphics[width=\hsize]{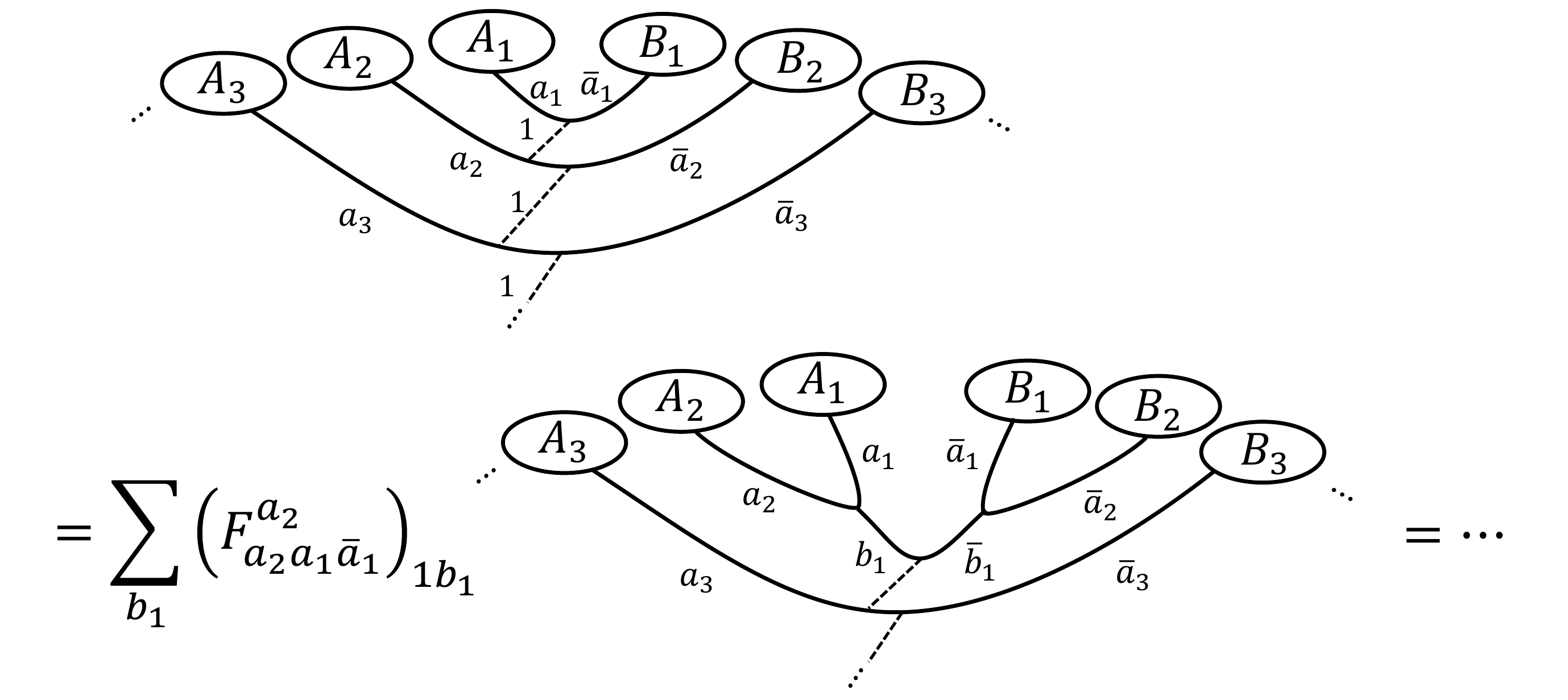}
\vspace{-10mm}
\end{center}
\caption{Illustration of the basis change using the F-matrix in order to decompose $\cH^1_{A^NB^N}$ into local parts A and B.}
\label{split}
\vspace{-5mm}
\end{figure}

We find that the AEE can be expressed in the following closed form. 
\bthm \label{Ainf}
For a pure state $\rho=|\psi\>\langle \psi|$ given by $|\psi\>=\sum_a\sqrt{p_a}|\psi_a\>$ with $|\psi_a \> \in \cH_A^a\ot \cH_B^{\bar a}$, it holds that
\begin{equation}\label{eq:Ainf}
E_A^\infty(\rho)=E_A^1(\rho)+\sum_ap_a\log d_a \, .
\end{equation}
\ethm
\begin{proof}
In order to derive Eq.~\eqref{eq:Ainf}, we have to express the $N$-copy state $\iota(|\psi\>^{\ot N})$  in a basis respecting the charges shared by $A$ and $B$, i.e.,~$\cH_{A^NB^N}^1=\bigoplus_c\cH_{A^N}^c\ot\cH_{B^N}^{\bar c}$ (see Fig.~\ref{split}). The $N$-copy state is given by 
\begin{align}\label{eq:ncopy}
\iota(|\psi\>^{\ot N})=\sum_{\bf a} \sqrt{p_{{\bf a}}}&|\psi_{a_1}\>\cdots|\psi_{a_N}\> \nonumber\\
&\ot|a_1{\bar a_1};1\>\cdots|a_N{\bar a_N};1\>\, ,
\end{align}
where $p_{\bf a}=p_{a_1}...p_{a_N}$ and $|a_i{\bar a_i};1\>$ denotes the basis vector of the fusion space $V_{a_i{\bar a_i}}^1$ given by Eq.~\eqref{def:fusionspace}. 
We perform a basis transformation on the state~\eqref{eq:ncopy}
in order to be able to split it into local parts $A$ and $B$. The explicit transformation is given by 
\begin{align}\label{Ntensor2}
|a_1{\bar a_1};1\>\cdots&|a_N{\bar a_N};1\>\nonumber\\
&= \sum_{{\bf b},c}\left(F^{a_2}_{a_2a_1{\bar a_1}}\right)_{1b_1}\left(F^{a_2}_{a_2b_1{\bar b_1}}\right)_{1b_2}\nonumber\\
&...\left(F^{a_N}_{a_Nb_{N-2}{\bar b_{N-2}}}\right)_{1c}
 |{\bf a},{\bf b},c\>|{\bf \bar a},{\bf \bar b},{\bar c}\> \, ,
\end{align} 
and by using the fact $(F_{ba{\bar a}}^b)_{1c}=\sqrt{\frac{d_c}{d_ad_b}}$ (see, e.g.,~\cite{2008AnPhy.323.2709B}), it is easy to see that 
\begin{align}  \label{eq:NcopyState}\hspace{-0.39cm}
\iota(|\psi\>^{\ot N})=&\sum_{{\bf a},{\bf b},c}\sqrt{\frac{p_{{\bf a}}d_c}{d_{{\bf a}}}}|\psi_{a_1}\>\cdots|\psi_{a_N}\>
|{\bf a},{\bf b},c\>|{\bf \bar a},{\bf \bar b},{\bar c}\>\,,\hspace{-1mm}
\end{align} 
where $d_{{\bf a}} =d_{a_1}...d_{a_N}$ and $|{\bf a},{\bf b},c\>=|a_1a_2;b_1\>\cdots|b_{N-2}a_N;c\>$ denotes an orthonormal basis of $V_{a_1...a_N}^c$ (similar for $|{\bf \bar a},{\bf \bar b},{\bar c}\>$ and $V_{{\bar a}_1...{\bar a}_N}^{\bar c}$). We denote the density matrix of the embedded state $\iota\left(|\psi\>^{\ot N}\right)$ by $\rho^N$. 
Taking the partial trace $\tau_B$ over the local part $B$, the reduced density matrix $\rho^N_A=\tau_B\left(\rho^N\right)$ is given by 
\begin{align}\label{rho^N}
\rho_A^N=\sum_{{\bf a},{\bf b},c}&\frac{p_{{\bf a}}d_c}{d_{{\bf a}}}\tr_B(|\psi_{a_1}\>\langle\psi_{a_1}|)\otimes \cdots\\
&\otimes\tr_B(|\psi_{a_N}\>\langle\psi_{a_N}|)\otimes|{\bf a},{\bf b},c\>\langle{\bf a},{\bf b},c|.
\end{align}
The reduced state $\rho_A^N$ can be regarded as a classical-quantum state, that is, a classical mixture of quantum states associated to $({\bf a},{\bf b},c)$ distributed according to $\frac{p_{{\bf a}}d_c}{d_{{\bf a}}}$. Using properties of the von Neumann entropy, it directly follows that
\begin{align}\label{E_A^N}
E_A^1\big(\iota(&|\psi\>^{\ot N})\big)=H\Big(\Big\{\frac{p_{\bf a}d_c}{d_{\bf a}}\Big\}\Big) +\nonumber \\ &
\sum_{{\bf a},{\bf b},c}\frac{p_{\bf a}d_c}{d_{\bf a}}\Big(E_A^1(|\psi_{a_1}\>)+\cdots
+E_A^1(|\psi_{a_N}\>)\Big)  \,.
\end{align}
The first term on the right hand side of  Eq.~\eqref{E_A^N} can be calculated as
\begin{align}\hspace{-0.45cm}
H\left(\left\{\frac{p_{\bf a}d_c}{d_{\bf a}}\right\}\right)&=-\sum_{{\bf a},{\bf b},c}\frac{p_{\bf a}d_c}{d_{\bf a}}\log \frac{p_{\bf a}d_c}{d_{\bf a}}\\
&=-\sum_{{\bf a},c}\frac{p_{\bf a}\dim V^c_{a_1...a_N} d_c}{d_{\bf a}}\log \frac{p_{\bf a}d_c}{d_{\bf a}}\label{eq:38}\\
&=-\sum_{{\bf a}}p_{\bf a}\Big(\log\frac{p_{a_1}}{d_{a_1}}+\cdots+\log\frac{p_{a_N}}{d_{a_N}}\Big)\nonumber
 \\& \quad - \sum_c\Big(\sum_{\bf a}\frac{p_{\bf a}\dim V_{a_1... a_N}^c d_c}{d_{\bf a}}\Big)\log d_c\label{eq:39}\\
&=N \Big(H(\{p_a\})+\sum_{a\in\cL}p_a\log d_a\Big)\nonumber\\
&\quad-\sum_{c\in\cL}q_c\log d_c\,,
\end{align}
with $q_c :=\sum_{\bf a}p_{\bf a}\dim V_{a_1\cdots a_N}^cd_c/d_{\bf a}$. 
For the second equality, we used the relationship $\sum_b 1 = \dim  V^c_{\bf a}$ for fixed charges ${\bf a}$ and $c$, which follows from~\eqref{dimfu}.  The third equality follows from Lemma~\ref{dimdim}. Note that Lemma~\ref{dimdim} implies that $q_c$ is a probability distribution, i.e., $q_c\geq0$ and $\sum_cq_c=1$.

The second term in Eq.~\eqref{E_A^N} can be simplified to $N\sum_{a}p_aE_A^1(|\psi_a\>)$.
Since $E_A^1(|\psi\>)=H(\{p_a\})+\sum_ap_aE_A^1(|\psi_a\>)$, we finally obtain Theorem~\ref{Ainf} by substituting each term in Eq.~\eqref{E_A^N} and calculating
\begin{align}
E_A^\infty(\rho) &=\lim_{N\to\infty}\frac{1}{N}E_A^1(\rho^N)\\
&=E_A^1(|\psi\>)+\sum_{a\in\cL}p_a\log d_a\,. 
\end{align}
\end{proof}
 In contrast to $E_A$ and $E_A^1$, $E_A^\infty$ depends not only on the Schmidt coefficients of the state, but also on the quantum dimensions of the anyon charges  shared by $A$ and $B$. 
As expected the additional contribution $\log d_a$ vanishes only if $a$ is abelian, i.e.,~$d_a=1$. Theorem~\ref{Ainf} indicates that $E_A^\infty$ coincides with the entanglement increase induced by simple anyonic excitations in 2-dimensional topologically ordered spin systems~\cite{PhysRevLett.111.220402} and conformal field theories~\cite{2014arXiv1403.0702H}. 
We further note that Eq.~\eqref{eq:Ainf} is similar to the EE defined by Hikami in the framework of topological quantum field theory~\cite{2008AnPhy.323.1729H}. 
This is particularly interesting since Hikami's EE and our $E_A^\infty$ follow from totally different approaches and motivations.  

\section{properties of AEE}\label{sec.5}
The AEE fulfills all necessary properties of an entanglement measure for anyonic bipartite pure states with total charge 1. In other words, $E_A^\infty$ is a non-negative function which is 0 if and only if the state is separable and it is non-increasing under anyonic LOCC operations. 
\bprop
For all $|\psi\> \in \bigoplus_{a}\cH_A^a\ot \cH_B^{\bar a}$, it holds that 
\begin{itemize}
\item[1)] AEE is a non-negative function
\begin{equation}
E_A^\infty(|\psi\>)\geq 0
\end{equation}
and the equality holds if and only if $|\psi\>$ is separable.
\item[2)] If $|\psi\>$ can be converted to $|\phi_j\>\in \bigoplus_{a}\cH_A^a\ot \cH_B^{\bar a}$ with probability $q_j$ by LOCC, then 
\begin{equation}
E^\infty_A(|\psi\>)\geq\sum_jq_jE^\infty_A(|\phi_j\>).
\end{equation}
\end{itemize}
\eprop
\begin{proof}
$1)$ The non-negativity follows directly from the definition. Moreover, by definition $E_A^1(|\psi\>)=0$ if $|\psi\>$ is separable. 
If the total charge is the vacuum, a separable state $|\psi\>$ is a state on $\cH_A^a\ot \cH_B^{\bar a}$ with $a$ and $\bar a$ abelian, such that $d_a$ is 1. 
Therefore, $E_A^\infty(|\psi\>)=0$ follows from Theorem~\ref{Ainf}. And for the converse, note that by Theorem~\ref{Ainf}, $E_A^\infty=0$ implies that $|\psi\>$ is separable and $d_a=1$.\\
$2)$
For any fixed $N$, $E_A^1(\rho^N)$ is computed by first embedding the state $\rho^N$ by $\cJ$ into a tensor product Hilbert space $\tilde \cH_{A_NB_N}$ and then evaluating standard entanglement entropy  of the embedded state. 
Under this embedding $\cJ$, any LOCC operation on the anyonic system is transformed to an LOCC operation on the bigger space $\cH_{A_NB_N}$. 
Since the entanglement entropy cannot increase under LOCC, neither can AEE. 
\end{proof}\noindent
Moreover, AEE satisfies the additivity for arbitrary two pure states.
\bprop
For any  two states $|\psi_1\>,|\psi_2\>$ with $|\psi_i\> \in \bigoplus_{a}\cH_{A_i}^a\ot \cH_{B_i}^{\bar a}$ holds that
\begin{equation}
E^\infty_{A_1A_2}\left(\iota(|\psi_1\>\ot|\psi_2\>)\right)=E_{A_1}^\infty(|\psi_1\>)+E_{A_2}^\infty(|\psi_2\>).
\label{fad}
\end{equation}
\eprop
\begin{proof}
Assume that $|\psi_i\>=\sum_a\sqrt{p^i_a}|\psi^i_a\>$ with $|\psi^i_a\> \in \cH_{A_i}^a\ot \cH_{B_i}^{\bar a}$ where $i=1,2$. The embedded state $\iota(|\psi_1\>\ot|\psi_2\>)$ 
is then given by
\begin{align}\label{eq30}
 \iota(|\psi_1\>\ot|\psi_2\>)=&\sum_{a_1,a_2,c}\sqrt{\frac{p_{a_1}^{1}p_{a_2}^{2}d_c}{d_{a_1}d_{a_2}}}|\psi^1_{a_1}\>|\psi^2_{a_2}\>\nonumber\\&\ot|a_2a_1;c\>|{\bar a_1}{\bar a_2};{\bar c}\> \, .
\end{align}
By using Theorem~\ref{Ainf}, we obtain that
\begin{align}
E_{A_1A_2}^\infty\left(\iota(|\psi_1\>\ot|\psi_2\>)\right)=&E_{A_1A_2}^1\left(\iota(|\psi_1\>\ot|\psi_2\>)\right)\nonumber\\&+\sum_{abc}\frac{p^1_ap^2_bd_c}{d_ad_b}\log d_c\,.
\label{fad1}
\end{align}
We can directly calculate the first term in Eq.~\eqref{fad1} through Eq.~\eqref{eq30} as 
\begin{align}
E_{A_1A_2}^1\left(\iota(|\psi_1\>\ot|\psi_2\>)\right) 
=&E_{A_1}^\infty(|\psi_1\>)+E_{A_2}^\infty(|\phi_2\>)\nonumber\\&-\sum_{abc}\frac{p^1_ap^2_bd_c}{d_ad_b}\log d_c \, .
\label{fad2}
\end{align}
By inserting Eq.~\eqref{fad2} into Eq.~\eqref{fad1} we see that additivity holds. 
\end{proof}

Let us now address the question which states maximize AEE. 
For qudit systems, the maximally entangled state (MES) $|\Psi_{\max}\>$ in a $d\times d$ system is a state with uniform Schmidt coefficients, i.e., $|\Psi_{\max}\>=\sum_{i=1}^d\frac{1}{\sqrt{d}}|{ii}\>$ for some product basis $\{|{ii}\>\}$ 
 and $E_A(|\Psi_{\max}\>)=\log d$.
However, since AEE also depends on the quantum dimensions, the anyonic MES maximizing AEE is different and has to follow a particular distribution over the superselection rules sectors. 
For simplicity, let us restrict to the situation where both systems $A$ and $B$  are given by $n$ anyons with charges $\{x_1,...,x_n\}$ and $\{{\bar x_1},...,{\bar x_n}\}$, where all $x_i$ are primitive.  
Then, the maximum of  $E_A^\infty$ over states in the corresponding fusion space is attained for 
\begin{equation} \label{eq:MaxEnt}
|\Psi^{{\bf x}}_{\max}\>=\sum_a\sqrt{\frac{\dim V_{x_1...x_n}^ad_a}{\prod_id_{x_i}}}|\psi_{\max}^a\>\,,
\end{equation}
where $|\psi_{\max}^a\>$ denotes the usual maximally entangled state in $\cH_A^a\ot \cH_B^{\bar a}$. Moreover, it holds that 
\begin{equation}
E_A^\infty(|\Psi^{{\bf x}}_{\max}\>)=\sum_i\log d_{x_i}.
 \end{equation}

To see this, let us set $D_{\bf x}^a= \dim V_{x_1...x_n}^a$ and $D_{x^N_i}^a= \dim V_{x_i... x_i}^a$. We first derive an upper bound on $E_A^\infty$ and then show that the 
state given by Eq.~\eqref{eq:MaxEnt} attains the upper bound. In order to calculate $E_A^1$ of an $N$-copy state $\rho^N$, we first embed it into a larger but unphysical Hilbert space which has a tensor product structure. The embedded $N$-copy state $\cJ(\rho^N)$ is a state on the non-physical Hilbert space $\cH_{A^N}\ot\cH_{B^N}\equiv(\bigoplus_d\cH_{A^N}^d)\ot(\bigoplus_d\cH_{B^N}^d)$.
The subspaces $\cH_{A^N}$ and $\cH_{B^N}$ contain $N$ $x_i$-anyons, and due to Eq.~\eqref{qudimn} we know that for sufficiently large $N$, $D_{x_i^N}^{b_i} \approx d_{x_i}^Nd_{b_i}/{\cal D}^2$. Hence, we have
\begin{align}
\dim \cH_{A^N} &=\dim \cH_{B^N} \nonumber\\&=\sum_a\sum_{b_1...b_n} \prod_iD_{x^N_i}^{b_i}\dim V_{b_1...b_n}^a \nonumber\\&
\approx \prod_id_{x_i}^N\sum_{a,\bf b}\frac{d_{b_1}...d_{b_n}}{{\cal D}^{2n}}\dim V_{b_1...b_n}^a \, ,
\end{align}
which implies that 
\begin{align}
\frac{1}{N}E_A^1(\iota(|\psi\>^{\ot N}))&\leq \frac{1}{N}\log \dim \cH_{A^N}\nonumber\\&=\frac{1}{N}\big(\sum_i\log d_{x_i}^N + O(1) \big) \, .
\label{upperbound}
\end{align}
Taking the limit $N$ to infinity in (\ref{upperbound}), we obtain the upper bound $E_A^\infty(|\psi\>)\leq \sum_i\log d_{x_i}$.

In order to see that $|\Psi^{\bf x}_{\max}\>$ attains the optimal value, we compute
\begin{align}
\nonumber E_A^\infty&(|\Psi^{{\bf x}}_{\max}\>)\nonumber\\=& H\left(\left\{\frac{D_{\bf x}^ad_a}{\prod_id_{x_i}}\right\}\right)\nonumber\\
&+\sum_a\frac{D_{\bf x}^ad_a}{\prod_id_{x_i}}\left(E_A^1(|\psi_{\max}^a\>)+\log d_a\right)\\
=&\sum_a\frac{D_{\bf x}^ad_a}{\prod_id_{x_i}}\left(-\log\frac{D_{\bf x}^ad_a}{\prod_id_{x_i}}+\log D_{\bf x}^a+\log d_a\right)\\
=&\sum_{a,j}\frac{D_{\bf x}^ad_a}{\prod_id_{x_i}}\log d_{x_j}=\sum_i\log d_{x_i}.
\end{align}
The last equality holds by Lemma~\ref{dimdim}. 
One example of the anyonic maximally entangled states is given if all pairs of anyons $x_i$ in A and $\bar x_i$ in B are created from the vacuum.

\section{Entanglement Distillation and Dilution}\label{sec.6}
In the following we provide operational meanings to AEE by relating it to the optimal asymptotic rates of entanglement distillation and entanglement dilution.
An entanglement distillation protocol converts copies of $|\psi\>$ to copies of the anyonic MES by LOCC operations. 
The rate of an entanglement distillation protocol is defined as the maximum numbers of distillable anyonic MESs per copy of $|\psi\>$ in the limit of infinite copies.  
The reverse task of generating copies of $\ket{\psi}$ from anyonic MES is called an entanglement dilution protocol. The rate is defined similarly to entanglement distillation.    
We define the distillable entanglement $E_D$ as the optimal rate of any entanglement distillation protocols and the entanglement cost $E_C$ as the optimal rate of any entanglement dilution protocols.

In order to ensure that the anyonic MES is well defined, we impose the same restrictions as used to specify the MES given by Eq.~\eqref{eq:MaxEnt}. Namely, we assume that system $A$ and $B$ are described by $n$ primitive anyons. 
We further assume that we can individually perform any special unitary on each sector by braiding with arbitrary accuracy. This is for instance satisfied for the Fibonacci anyon~\cite{Freedman2002}. 
We then obtain the equivalence of AEE with $E_D$ and $E_C$.  
\bthm \label{thm:distillcost}
For  any pure bipartite state $|\psi\>=\sum_a\sqrt{p_a}|\psi_a\>$ 
with a primitive charge $a\in\cL$ satisfying $p_a \neq 0$, it holds that
\begin{equation}
E_D(|\psi\>)=\frac{E^\infty_A(|\psi\>)}{E^\infty_A(|\Psi_{\max}^{\bf x}\>)}= E_C(|\psi\>).
\end{equation}
\ethm 

In the following, we provide a sketch of the proof including the essential ideas. The  technical details of the proof are carried out in Appendix~\ref{sec:Dist}.

We first discuss the optimal entanglement distillation protocol and show achievability. 
Inserting  the Schmidt decomposition $|\psi_a\>=\sum_{i_a} \sqrt{\lambda_{i_a} }\ket {i_a}\ket {i_a}$ in Eq.~\eqref{eq:NcopyState}, 
we see that $\iota(|\psi\>^{\ot N})$ can be written as 
\begin{equation}
\iota(|\psi\>^{\ot N})=\sum_{{\bf a},{\bf b},c}\sqrt{ \frac{p_{{\bf a}}\lambda_{{\bf i_a}} d_c}{d_{{\bf a}}}}
|{\bf a},{\bf b},c,{\bf i_a}\>|{\bf \bar a},{\bf \bar b},{\bar c},{\bf i_a}\> \,  ,
\end{equation}
where $\lambda_{{\bf i_a}} = \lambda_{i_{a_1}}...\lambda_{i_{a_N}}$. Note that the amplitudes $\frac{p_{{\bf a}}\lambda_{{\bf i_a}}}{d_{{\bf a}}}$ follow an identical and independent distribution according to $p_a\lambda_{i_a}$ with weight $1/d_a$. We can project for every $c$ onto the $\delta$-typical subspace corresponding to $p_a\lambda_{i_a}$ and bring the success probability arbitrary close to $1$ by choosing a sufficiently large $N$. 

In a next step, we project onto the different type-classes in the $\delta$-typical subspace such that all the amplitudes for fixed $c$ are equal.  
Using the standard typicality properties~\cite{CoverThomas199108,Wilde201304}, we know that the dimension of the resulting typical subspace is about $2^{N H(\{p_a \lambda_{i_a}\})}=2^{N E^1_A(\ket \psi)}$ for any $c$. Moreover, the additional fusion dimensions labeled by ${\bf b}$ count up for given $c$ to $\dim V^c_{ {\bf a_t}}$ where ${\bf a_t}$ is a typical sequence. Using the scaling behavior of the fusion space and the typicality of ${\bf a_t}$, we find that $\dim V^c_{ {\bf a_t}} \approx (d_c/\cD^2)\Pi_a d_a^{Np_a}$ (see Appendix~\ref{sec:Dist}). Hence, the total dimension of the non-trivial subspace in sector $c$ is about $(d_c/\cD^2) 2^{N H(\{p_a \lambda_{i_a}\})}\dim V^c_{ {\bf a_t}} \approx (d_c/\cD^2) 2^{N (E^1_A(\ket \psi) + \sum_a p_a \log d_a)} = (d_c/\cD^2) 2^{N E^\infty_A(\ket \psi)}$. 

Note now that on each sector $c$, $\iota(|\Psi^{{ x^{n}}}_{\max}\>^{\otimes L})$ corresponds to the MES in the fusion space $V_{x^{nL}}^c$ with dimension $\dim V_{x^{nL}}^c = (d_c/\cD^2)d_{x}^{nL}$. Hence, the dimensions of the subspaces match for every $c$ if we choose $L\approx N E^\infty_A(\ket \psi)/\log d_x^n = N E^\infty_A(\ket \psi)/E^\infty_A(|\Psi_{\max}^{\bf x}\>)$. Therefore, a simple local basis transformation  converts the projected state into $L$ anyonic MESs.  This concludes the achievability for entanglement distillation.

The optimal entanglement dilution protocol is based on LOCC convertibility of each individual sector. For that we use similar methods as for the distillation protocol to show that, on each sector, the amplitudes of the typical part of $N$ copies of $|\psi\>$ are majorized by those of about $L=N E^\infty_A(\ket \psi)/E^\infty_A(|\Psi_{\max}^{\bf x}\>)$ copies of $|\Psi_{\max}^{\bf x}\>$. 
Optimality of the two protocols can be argued by concatenating entanglement distillation and dilution.\\

\section{Conclusion}\label{sec.7}
We have presented an operational approach to bipartite entanglement of pure anyon chains by introducing the entanglement measure AEE. We have showed that AEE is equal to the contribution of anyonic excitations in topologically ordered phases to TEE ~\cite{2008AnPhy.323.1729H, 2008JHEP...05..016D, PhysRevLett.111.220402, 2014arXiv1403.0702H} and further showed that it characterizes the optimal entanglement distillation and dilution rates. These results provide  operational meanings to the TEE of anyonic excitations, and moreover, identify  the TEE as the operationally accessible entanglement for fault-tolerant quantum information processing. 

We point out that our result may be applied to other interesting situations obeying similar superselection rules like anyons, e.g., angular momenta with no shared reference frame~\cite{PhysRevLett.91.027901}. Moreover, it would be also desirable to extend  AEE to more general splittings of anyonic systems as considered in Ref.~\cite{PhysRevB.89.035105}. 

\section*{Acknowledgment}
This work is supported by ALPS, 
the Project for Developing Innovation Systems
of MEXT, Japan, and JSPS KAKENHI
(Grant No. 23540463, No. 23240001 and No. 26330006).
We also gratefully acknowledge the ELC project (Grant-in-Aid for
Scientific Research on Innovative Areas MEXT KAKENHI (Grant No.
24106009)) for encouraging the research presented in this letter. 
FF acknowledges support from Japan Society for the Promotion of Science (JSPS) by KAKENHI grant No. 24-02793.
 
\appendix
\section{Proof of Theorem 2: Optimal Entanglement Distillation and Dilution Protocol}\label{sec:Dist}
\subsection{Entanglement Distillation}
In the following, we first present a distillation protocol and show that its rate is given by AEE. Recall first that both systems $A$ and $B$ are assumed to be given by chains of anyons with primitive charges 
${\bf x }= \{x_1,...,x_n\}$ and ${\bf\bar x}=\{\bar x_1,...,x_n\}$. 
In order to simplify notations, we further require that $x=x_1=\cdots=x_n$ which can be straightforwardly generalized  to arbitrary primitive charges. 
Moreover, we assume that we can perform arbitrary unitary operations in $SU(N)$ on each sector $\cH_A^a$ (or $\cH_B^b$) by only braiding operations, i.e., we can perform universal topological quantum computation. We consider two parties Alice and Bob who holds anyonic systems $A$ and $B$, respectively. 
At the beginning of the protocol, Alice and Bob share many identical copies of a state 
$|\psi\>= \sum_a\sqrt{p_a}|\psi_a\>$ with $|\psi_a\> \in \cH_{A}^a\ot \cH_{B}^{\bar a}$. 
Using the Schmidt decomposition, each $|\psi_{a_j}\>$ can be written as  
\begin{equation}
|\psi_{a_j}\>=\sum_{i_{a_j}}\sqrt{\lambda_{i_{a_j}}}|i_{a_j}\>|i_{\bar a_j}\>,
\end{equation}
where $\{\lambda_{i_{a_j}}\}$ denotes the Schmidt coefficients of $|\psi_{a_j}\>$. 
According to~\eqref{eq:NcopyState}, the N-copy state $|\psi^N\>\equiv\iota(|\psi\>^{\ot N})$ can be written as
\begin{equation}\label{ncopy}
|\psi^N\>=\sum_{{\bf a,b_a^c},c,{\bf i_a}}\sqrt{\frac{p_{\bf a}d_c}{d_{\bf a}}\lambda_{{\bf i_a}}}
|{\bf a,b_a^c},c,{\bf i_a}\>_{A^N}|{\bf {\bar a},{\bar b_a^c}},{\bar c},{\bf i_{\bar a}}\>_{B^N}\,,
\end{equation} 
where ${\bf i_a}=(i_{a_1},...,i_{a_N})$ and $\lambda_{\bf i_a}=\lambda_{i_{a_1}}...\lambda_{a_{i_N}}$. Here we explicitly denote the dependence of $\bf b$ and the subsystems where the vectors belongs to.
Our goal is to obtain $L$ copies of the maximally entangled state 
\begin{align}
\iota(|\Psi^{\bf x}_{\max}\>&^{\ot L})= \sum_c\sqrt{\frac{\dim V^c_{x^{n L}}d_c}{(d_x)^{nL}}}\nonumber\\
&\times\left[\frac{1}{\sqrt{\dim V_{x^{nL}}^c}}\sum_{y^c}^{\dim V_{x^{nL}}^c}|y^c\>_{A^N}|y^{\bar c}\>_{B^N}\right] \,,\label{max state}
\end{align}
with maximal $L$.

Let us focus on the $\delta$-strongly typical set $T_\delta^N$ induced by the distribution $p_a\lambda_{i_a}$ which is defined as
\begin{equation}\label{def:typ}
T_\delta^N := \Big\{({\bf a,i_a}):\big|\frac{1}{N}N(a,i_a|{\bf a,i_a})-p_a\lambda_{i_a}\big|\leq\delta \Big\}\,,
\end{equation} 
where $N(a,i_a|{\bf a,i_a})$ is the number of $(a,i_a)$ in the sequence ${\bf a,i_a}$. 
Performing the projective measurement on the corresponding typical subspace, we obtain the state
\begin{align}\label{typical state}
|\psi^N_{\rm typ}\>=\frac{1}{\sqrt{P_\delta^N}}&\sum_{c}\sum_{{\bf a,i_a}\in T_\delta^N}\sum_{\bf b_a^c}\sqrt{\frac{p_{\bf a}d_c}{d_{\bf a}}\lambda_{\bf i_a}}\nonumber\\&\times
|{\bf a,b_a^c},c,{\bf i_a}\>_{A^N}|{\bf {\bar a},{\bar b_a^c}},{\bar c},{\bf i_{\bar a}}\>_{B^N}\, ,
\end{align}
where for any $\epsilon>0$, we can choose $N$ large enough so that the success probability $P_\delta^N$ is at least $1-\epsilon $. This is guaranteed by the typicality properties for independently and identically distributed variables (see, e.g., Ref.~\cite{CoverThomas199108}).   
Next, we consider the type class $T_t^N$ which is defined by
\begin{equation}\label{def:type}
T_t^N := \left\{({\bf a,i_a}):\frac{1}{N}N(a,i_a|{\bf a,i_a})=t^{a,i_a}\right\}\,,
\end{equation} 
where $t^{a,i_a}$ is the $(a,i_a)$ component of the probability distribution $t$. 
By defining the set of types in $T_\delta^N$ by
\begin{equation}
\tau_\delta=\left\{t:\left|t^{a,i_a}-p_a\lambda_{i_a}\right|\leq\delta\right\}\,,
\end{equation}
 $T_\delta^N$ can be decomposed as
\begin{equation}
T_\delta^N=\bigcup_{t\in\tau_\delta}T_t^N\,.
\end{equation}
The cardinality of the set $\tau_\delta$ is bounded by $|\tau_\delta|<(N+1)^d$, where $d=\sum_a\dim V_{x^n}^a$. 
Expanding into the different type classes, $|\psi_{\rm typ}^N\>$ can be written as 
\begin{widetext}
\begin{align}\label{type state}
|\psi_{\rm typ}^N\>& =  \frac{1}{\sqrt{P_\delta^N}}\sum_{t\in\tau_\delta}\sum_{c}\sum_{{\bf a,i_a}\in T_t^N}\sum_{\bf b_a^c}\sqrt{\frac{p_{\bf a}d_c}{d_{\bf a}}\lambda_{\bf i_{a}}}|{\bf a,b_a^c},c,{\bf i_a}\>_{A^N}|{\bf {\bar a},{\bar b_a^c}},{\bar c},{\bf i_{\bar a}}\>_{B^N}
\\ & =  \frac{1}{\sqrt{P_\delta^N}}
\sum_{t\in\tau_\delta}\sqrt{\frac{\prod_{a,i_a}(p_a\lambda_{i_a})^{Nt^{a,i_a}}}{\prod_ad_a^{Nt^a}}}\sum_{c}\sqrt{d_c}\sum_{{\bf a,i_a}\in T_t^N}\sum_{\bf b_a^c}|{\bf a,b_a^c},c,{\bf i_a}\>_{A^N}|{\bf {\bar a},{\bar b_a^c}},{\bar c},{\bf i_{\bar a}}\>_{B^N}\,,
\end{align}
\end{widetext}
where $t^a=\sum_{i_a}t^{a,i_a}\,$.

In the next step of the protocol, we perform a measurement of type $t$ and obtain the state 
\begin{align}\label{t-state}
\sqrt{\frac{\prod_{a,i_a}(p_a\lambda_{i_a})^{Nt^{a,i_a}}}{P_\delta^N q_t^N\prod_ad_a^{Nt^a}}}&\sum_{c}\sqrt{d_c}\nonumber
\\\times\sum_{{\bf a,i_a}\in T_t^N}\sum_{\bf b_a^c}&|{\bf a,b_a^c},c,{\bf i_a}\>_{A^N}|{\bf {\bar a},{\bar b_a^c}},{\bar c},{\bf i_{\bar a}}\>_{B^N}\,,
\end{align}
with probability $q_t^N$. The probability $q_t^N$ is given by
\begin{equation}
q_t^N=\frac{|T_\delta^N|\prod_{a,i_a}(p_a\lambda_{i_a})^{Nt^{a,i_a}}}{P_\delta^N}
\end{equation}
and bounded by~\cite{CoverThomas199108}
\begin{equation}
(N+1)^{-d}2^{-ND(t\|p\lambda)}\leq q_t^N\leq2^{-ND(t\|p\lambda)}\,,
\end{equation}
where $D(t\|p\lambda)$ is relative entropy of $t$ and $p\lambda$.

In the following, we denote  the dimension of the subspace corresponding to type $t$ in sector $c$ by 
\begin{equation}
N_t^c=\sum_{{\bf a,i_a}\in T_t^N}\sum_{\bf b_a^c}=|T_t^N|\dim V_{\bf a^t}^c\,.
\end{equation}
For every $c$ and fixed $t$, we relabel $({\bf a}, {\bf i_a},{\bf b_a^c})$ by $\alpha^c\in\{1,...,N_t^c\}$. 
Using this notations, we write the state given in Eq.~\eqref{t-state} as
\begin{align}
&\sqrt{\frac{\prod_{a,i_a}(p_a\lambda_{i_a})^{Nt^{a,i_a}}}{P_\delta^Nq_t^N\prod_ad_a^{Nt^a}}}\sum_{c}\sqrt{d_c}\sum_{\alpha^c=1}^{N_t^c}|\alpha^c\>_{A^N}|\alpha^{\bar c}\>_{B^N}
\\=&\sqrt{\frac{|T_t^N|\prod_{a,i_a}(p_a\lambda_{i_a})^{Nt^{a,i_a}}}{P_\delta^Nq_t^N}}\sum_{c}\sqrt{\frac{d_c\dim V^c_{\bf a^t}}{\prod_ad_a^{Nt^a}}}\nonumber
\\&\quad\times\sum_{\alpha^c=1}^{N_t^c}\frac{1}{\sqrt{N_t^c}}|\alpha^c\>_{A^N}|\alpha^{\bar c}\>_{B^N}
\\=&\frac{1}{\sqrt{P_\delta^N}}\sum_{c}\sqrt{\frac{d_c\dim V^c_{\bf a^t}}{\prod_ad_a^{Nt^a}}}\sum_{\alpha^c=1}^{N_t^c}\frac{1}{\sqrt{N_t^c}}|\alpha^c\>_{A^N}|\alpha^{\bar c}\>_{B^N}\,.
\end{align}

As shown in Ref.~\cite{Wilde201304}, for all $t\in\tau_\delta$ we can bound
\begin{equation}
|T_t^N|\geq 2^{N[H(p\lambda)-\eta(d\delta)-\frac{d}{N}\log(N+1)]}\,,
\end{equation}
where $\eta(d\delta)$ is a function such that $\eta(d\delta)\to0\;(\delta\to 0)$. 
Therefore, using Eq.~\eqref{qudimnprimitive}, we obtain a lower bound on $N_t^c$ via
\begin{align}
N_t^c&= |T_t^N|\dim V_{\bf a^t}^c\\
&\geq2^{N[H(p\lambda)-\eta(d\delta)-\frac{d}{N}\log(N+1)]}
\nonumber\\
&\quad\times\frac{\prod_{a\in {\cal L}}d_a^{N(p_a-\delta)}d_c}{{\cal D}^2}(1+{\cal O}(v_c^{Np_a}))\\
&=2^{N(H(p\lambda)+\sum_ap_a\log d_a-\eta(d\delta)-\frac{d}{N}\log(N+1)-\delta\sum_a\log d_a)}
\nonumber\\
&\quad\times\frac{d_c}{{\cal D}^2}(1+{\cal O}({v'}_c^N))\,,
\end{align}
where $|v_c|,|v'_c|<1$ for all $c\in{\cal L}$. 
Note that to use Eq.~\eqref{qudimnprimitive}, we have to make the assumption that $p_a\neq0$ for at least one primitive charge $a$.  
Let us set 
\begin{align}
M^c =& 2^{N(H(p\lambda)+\sum_ap_a\log d_a-\eta(d\delta)-\delta\sum_a\log d_a)}\nonumber\\
&\times2^{N(-\frac{d+1}{N}\log(N+1)-\frac{\chi_{N}}{N}\log d_x^n)}\frac{d_c}{{\cal D}^2}\,,
\end{align}
where $\chi_N$ is a constant such that $0\leq\chi_N<1$.
Then, we divide each $N_t^c$ dimensional sector into $\lfloor \frac{N_t^c}{M^c}\rfloor$ orthogonal subspaces of dimension $M^c$ and the rest. 
Since the dimension of the rest subspace is strictly smaller than $M^c$, the probability that the projection onto the $M^c$-dimensional subspaces fails is less than $\frac{M^c}{N_t^c}$. 
This error probability is bounded by
\begin{equation}
\frac{M^c}{N_t^c}\leq2^{-\log(N+1)-\chi_N\log d_x^n}\,.
\end{equation}
Hence, by measuring in which orthogonal subspaces the state falls, we obtain a state that is unitary equivalent to 
\begin{equation}
|\tilde \psi^N\>=\frac{1}{\sqrt{P_\delta^N}}\sum_{c}\sqrt{\frac{d_c\dim V^c_{\bf a^t}}{\prod_ad_a^{Nt^a}}}\sum_{y^c=1}^{M^c}\frac{1}{\sqrt{M^c}}|y^c\>_{A^N}|y^{\bar c}\>_{B^N}\,.\label{a24}
\end{equation}
with probability at least 
\begin{equation}
1-2^{-\log(N+1)-\chi_N\log d_x^n}\,.
\end{equation}

Recalling the form of $L$ copies of the maximally entangled state in Eq.~\eqref{max state}, we see  that Eq.~\eqref{a24} has a similar form, such that it remains to confirm that the amplitudes match.

For that we consider a set $\{L^c\}$ such that for all $c\in{\cal L}$, $\dim V_{x^{nL}}^c=M^c$. 
Then, since $\dim V_{x^{nL}}^c=\frac{d_x^{nL}d_c}{{\cal D}^2}(1+{\cal O}(w_c^L ))$ with $|w_c|<1$, we have that 

\begin{align}
L^c=&\frac{1}{\log d_x^n}N\{E_A^\infty(|\psi\>)-\eta(d\delta)-\frac{d+1}{N}\log(N+1)\nonumber\\
&-\delta\sum_a\log d_a)-\log(1+{\cal O}(w^N_c))\}-\chi_N\,.
\end{align}
Then, we set the number of copies $L$ as
\begin{align}\label{distL}
L=&\frac{1}{\log d_x^n}N\{E_A^\infty(|\psi\>)-\eta(d\delta)-\frac{d+1}{N}\log(N+1)\nonumber\\
&-\delta\sum_a\log d_a)-\log(1+K|w^N|)\}-\chi_N\,,
\end{align}
where we choose $0<K<\infty$ and $w$ to bound the function ${\cal O}(w_c^N)$ by $|{\cal O}(w_c^N)|\leq K |w^N|$ for all $c$ and large enough $N$. 
Thus $L\leq L^c$ and $\dim V^c_{x^{nL}}\leq M^c$ for all $c$ in this setting. 
Note that we can choose the constant $0\leq\chi_N<1$ to make $L$ a natural number beforehand.

In order to see that $|\tilde\psi^N\>$ converges to the maximally entangled state $\iota(|\Psi^{\bf x}_{\max}\>^{\ot L})$, we consider an inner product between $|\tilde\psi^N\>$ and $\iota(|\Psi^{\bf x}_{\max}\>^{\ot L})$ given by
\begin{align*}
&\langle\tilde\psi^N|\iota(|\Psi^{\bf x}_{\max}\>^{\ot L}\>\\&=\sum_c\sqrt{\frac{d_c^2\dim V_{x^{nL}}^c\dim V_{\bf a^t}^c}{P_\delta^Nd_x^{nL}\prod_ad_a^{Nt^a}}}
\frac{\dim V_{x^{nL}}^c}{\sqrt{\dim V_{x^{nL}}^cM^c}}\\
&=\sum_cd_c\sqrt{\frac{d_c^2}{{\cal D}^4}(1+{\cal O}(w_c^L))(1+{\cal O}(z_c^N))}\sqrt{\frac{\dim V_{x^{nL}}^c}{M^c}}\\
&=\sum_c\frac{d_c^2}{{\cal D}^2}\sqrt{(1+{\cal O}(w_c^L))(1+{\cal O}(z_c^N))}\sqrt{\frac{1+{\cal O}(w_c^L)}{1+K|w|^N}}\,.
\end{align*}
Here we used the relation $\dim V_{\bf a^t}^c=\frac{\prod_ad_a^{Nt^a}d_c}{{\cal D}^2}(1+{\cal O}(z^N_c))$, where $|z_c|<1$. 
Since $|w_c|, |z_c|, |w|<1$, we find that the absolute value of the inner product converges to 1 for $N\to\infty$. Therefore, the fidelity between two states $F(|\tilde\psi^N\>,\iota(|\Psi^{\bf x}_{\max}\>^{\ot L}\>)$ also converges to 1.   
The asymptotic rate of this distillation protocol is given by 
\begin{equation}
\frac{L}{N}\xrightarrow{N\to \infty,\delta \to 0}\frac{E_A^\infty(|\psi\>)}{E_A^\infty(|\Psi_{\max}^{\bf x}\>)}\,
\end{equation}
and the success probability is 
$(1-\epsilon )(1-2^{-\log(N+1)-\chi_N\log d_x^n})$, which converges to 1 for $N\to\infty$ and $\epsilon \to 0$.
\subsection{Entanglement Dilution}
Standard entanglement dilution protocols for qubits are based on quantum teleportation~\cite{PhysRevA.53.2046}. 
However, quantum teleportation of anyonic systems has not been established yet. 
In the following, we derive the rate of a dilution protocol without using teleportation but instead,  majorization and LOCC convertibility. 

We start from $L$ copies of the maximally entangled state 
\begin{align}
\iota(|\Psi^{\bf x}_{\max}\>^{\ot L})&= \sum_c\sqrt{\frac{\dim V^c_{x^{n L}}d_c}{(d_x)^{(nL)}}}\nonumber\\
&\quad\times\frac{1}{\sqrt{\dim V_{x^{nL}}^c}}\sum_{y^c=1}^{\dim V_{x^{nL}}^c}|y^c\>_{A^N}|y^c\>_{B^N}.
\end{align}
Our goal is to produce $N$ copies of $|\psi\>$ by using only LOCC. Let us consider the normalized state 
obtained by the projection onto the typical subspace corresponding to $T_\delta^N$ using  Eq.~\eqref{typical state}
\begin{align}
|\psi_{\rm typ}^N\>&=\sum_c\sqrt{Q^c}\sum_{{\bf a,i_a}\in T_\delta^N}\sum_{\bf b_a^c}\sqrt{\frac{p_{\bf a}d_c}{P_\delta^NQ^cd_{\bf a}\lambda_{\bf i_a}}}
\nonumber\\
&\quad\times|{\bf a,b_a^c},c,{\bf i_a}\>_{A^N}|{\bf {\bar a},{\bar b_a^c}},{\bar c},{\bf i_{\bar a}}\>_{B^N}\,,
\end{align}
where $Q^c=\sum_{{\bf a,i_a}\in T_\delta^N}\sum_{\bf b_a^c}\frac{p_{\bf a}d_c}{P_\delta^Nd_{\bf a}}\lambda_{\bf i_a}$. 
The Schmidt rank of each sector $c$ is bounded by 
\begin{align*}
\sum_{{\bf a,i_a}\in T_\delta^N}\sum_{\bf b_a^c} 1 &=
\sum_{{\bf a,i_a}\in T_\delta^N}\dim V_{\bf a}^c\\
&\leq\sum_{{\bf a\rq{},i_a}\in T_\delta^N}\frac{\prod_{a\rq{}}d_{a\rq{}}^{Np_a+\delta}d_c}{{\cal D}^2}(1+{\cal O}(w_c^N))\\
&=|T_\delta^N|\frac{\prod_{a}d_{a}^{Np_a+\delta}d_c}{{\cal D}^2}(1+{\cal O}(w_c^N))\,  , \label{eq75}
\end{align*}
where the first inequality is due to Eq.~\eqref{qudimnprimitive} and the fact that $(\bf a,i_a)\in T_\delta^N$. 
Since $|T_\delta^N|$ can be bounded by $2^{N(H(p\lambda)+\delta)}$ (see, e.g., Ref.~\cite{CoverThomas199108}), we can further bound
\begin{align}
\sum_{{\bf a,i_a}\in T_\delta^N}\sum_{\bf b_a^c} &\leq2^{N(H(p\lambda)+\delta)}2^{N\left(\sum_a(p_a+\delta)\log d_a\right)}\frac{d_c}{{\cal D}^2}(1+{\cal O}(w_c^N))\\
&=2^{N\left(E_A^\infty(|\psi\>)+\delta(1+\sum_a\log d_a)\right)}\frac{d_c}{{\cal D}^2}(1+{\cal O}(w_c^N))\,\label{eq79}.
\end{align}

If we choose $L$ as 
\begin{align}\label{diluteL}
L=\left\lfloor\frac{N\left(E_A^\infty(|\psi\>)+\delta(1+\sum_a\log d_a)\right)}{\log d_x^n}+1\right\rfloor\,,
\end{align}
we find by using Eq.~\eqref{eq79} for large $N$ that 
\begin{align*}
\dim V_{x^{nL}}^c=\frac{d_c2^{L\log d_x^n}}{\cD^2}(1+\cO(z_c^N))\geq\sum_{{\bf a,i_a}\in T_\delta^N}\dim V_{\bf a}^c\,.
\end{align*}
Therefore, for all $c\in {\cal L}$ and large $N$, the majorization relation
\begin{equation}
\left\{\frac{1}{\dim V^c_{x^{nL}}}\right\}\preceq \left\{ \frac{p_{\bf a}d_c}{P_\delta^NQ^cd_{\bf a}}\lambda_{\bf i_a}\right\}\,
\end{equation}
holds.  Here, we have extended trivially the size of the domain of the probability distribution on the right hand side to match the left hand side. 
By using the LOCC convertibility theorem under superselection rules~\cite{PhysRevLett.92.087904}, $L$ copies of the maximally entangled state can be deterministically converted to $|\psi_{typ}^N\>$. 

There is no error in this protocol and for any $\epsilon>0$ and we obtain  
\begin{align}
|\tilde\psi^N\>&=\sum_c\sqrt{\frac{\dim V^c_{x^{n L}}d_c}{(d_x)^{(nL)}}}\sum_{{\bf a,i_a}\in T_\delta^N}\sum_{\bf b_a^c}\sqrt{\frac{p_{\bf a}d_c}{P_\delta^NQ^cd_{\bf a}}\lambda_{\bf i_a}}\nonumber\\&\quad\times
|{\bf a,b_a^c},c,{\bf i_a}\>_{A^N}|{\bf {\bar a},{\bar b_a^c}},{\bar c},{\bf i_{\bar a}}\>_{B^N}\,.
\end{align}
By a simple calculation, the fidelity between $|\tilde\psi^N\>$ and $|\psi^N\>$ can be bounded by
\begin{align}\hspace{-0.8cm}
F( |\tilde\psi^N\>,|\psi^N\>) \geq&(1-\epsilon)^2(1-K'|z^N|)(1-K|w^N|)\nonumber\\&\xrightarrow{N\to\infty,\epsilon\to0} 1\,.
\end{align}
The asymptotic rate of this protocol is given as claimed by 
\begin{equation}
\frac{L}{N}\xrightarrow{N\to \infty,\delta\to0}\frac{E_A^\infty(|\psi\>)}{E_A^\infty(|\Psi_{\max}^{\bf x}\>)}\,.
\end{equation}

\subsection{Optimality}
Finally, we show that the asymptotic rates of the distillation and dilution protocol presented in the previous two sections are optimal. 
The argument is the same as to show optimality for the qubit case~\cite{PhysRevA.53.2046}. 
Let us replace $L$ in Eq.~\eqref{distL} by $L_D$ and in Eq.~\eqref{diluteL} by $L_C$. 
In the distillation protocol, 
we obtain $L_D$ copies of the maximally entangled state from $N$ copies of $|\psi\>$ (with small errors). 
Let us assume that there exists 
a distillation protocol which performs strictly better than the protocol and obtains $\lfloor L_D+\xi N \rfloor (\xi>0)$ copies of the maximally entangled state from $N$ copies of $|\psi\>$. By using the  dilution protocol presented in the previous subsection, we obtain at least
$N'$ copies of $|\psi\>$, where $N'$ is given by
\begin{equation}
N'=\left\lfloor\frac{\lfloor L_D+\xi N\rfloor}{L_C}N\right\rfloor\,.
\end{equation}
Therefore, we have
\begin{equation}
\lim_{N\to\infty}\frac{N'}{N}\geq\frac{E_A^\infty(|\psi\>)+\xi\log d_x^n}{E_A^\infty(|\psi\>)}>1\,,
\end{equation}
and thus, in the limit of $N\to \infty,\delta,\epsilon\to0$, the amount of entanglement can be increased by LOCC. 
This conflicts the property of LOCC, which implies that such a protocol does not exist.  
Therefore the obtained asymptotic rate of the distillation protocol is optimal. Similarly, one can prove that the asymptotic rate of the dilution protocol 
is optimal.

\section{Non-Abelian Total Charge and Anyonic Purification}\label{sec:nonvac}

In the main text, we restricted ourselves to a system with total charge $1$ of which the Hilbert space can be written as $\cH^1=\bigoplus_{a \in {\cal L}}\cH_A^a\ot \cH_B^{\bar a}$.
Let us provide an outlook what happens if we relax this condition and allow a general system with total charge $c$. In this situation, the Hilbert space is given by
\begin{equation}
\cH^c=\bigoplus_{a,b\in {\cal L}}\cH_A^a\ot \cH_B^b\ot V_{ab}^c,\label{genhil}
\end{equation}
where $\dim V_{ab}^c$ can be strictly larger than $1$.  There are several problems arising in this situation, among them the impossibility to generally define a state on the combination of two systems with charge not equal $1$ by knowing the state only on partial systems. Or physically, a joint preparation of the combined system is required since the splitting into subsystems with charge $c$ and $\bar c$ does not have a tensor product structure, and thus, does not allow local preparations.  

\begin{figure}[htbp]\vspace{1cm}
\includegraphics[width=1\hsize]{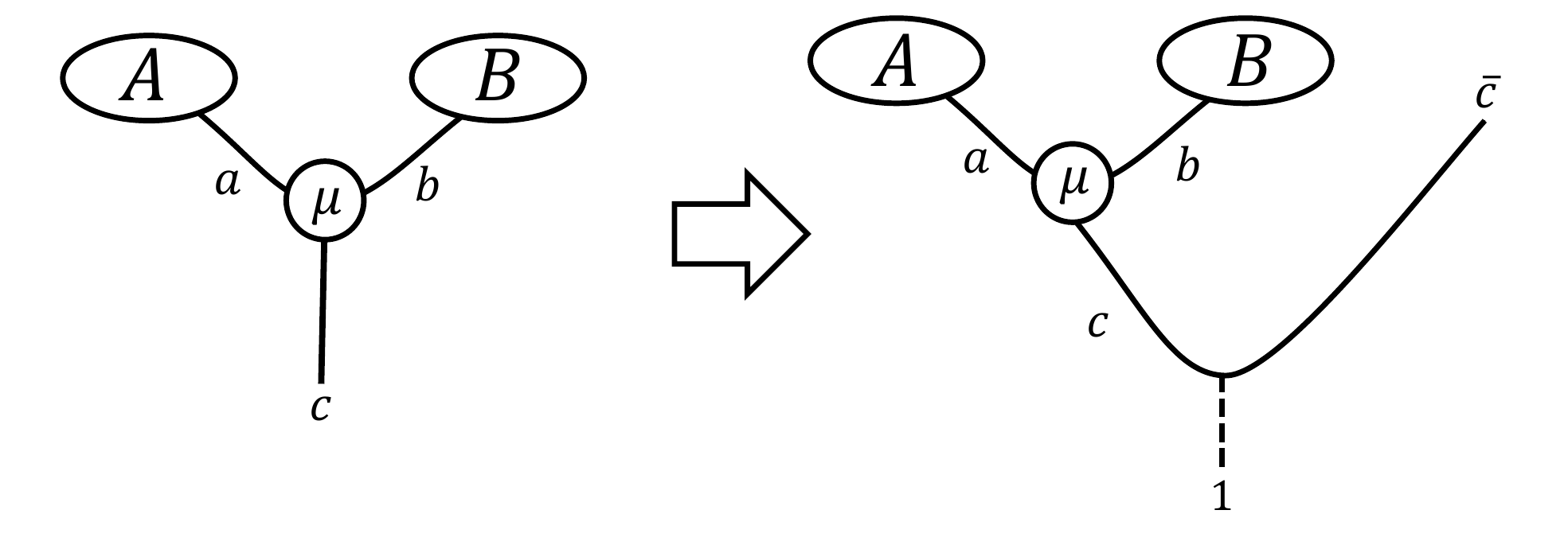}
\caption{The right hand side illustrates the anyonic purification of the system on the left hand side with total charge $c$.}
\label{puri}
\end{figure}

In order to deal with this problems, it is convenient to complement the system by introducing a reference system with total charge given by the anti-charge $\bar c$ such that the combined system has  charge $1$. We call this extension of the system the anyonic purification and it leads to a total Hilbert space 
\begin{equation}
\tilde \cH^1=\bigoplus_{a,b\in {\cal L}}\cH_A^a\ot \cH_B^b\ot V_{ab}^c\ot V_{c\bar c}^1
\end{equation}
including the $1$-dimensional Hilbert space $V_{c\bar c}^1$. A similar method to treat non-trivial total charges has already been discussed in Ref.~\cite{kitaev2004superselection}.
Graphically, the anyonic purification is illustrated in Fig.~\ref{puri}. Note that since $V_{c\bar c}^1$ is only one-dimensional, every state on the system with total charge $c$ allows a unique extension to the purified system $\tilde \cH^1$ up to a global phase which can be neglected. 
 
By using this purified system the total charge is now guaranteed to be $1$. Thus, two purified systems can be combined in the same way as described in the main text and the joint state is uniquely defined. Note that this construction corresponds exactly to the requirement that the two states are independently prepared. From an operational perspective, this local preparation can be only achieved by the creation of a state from vacuum and then discarding a part of it. However, the operations and manipulations on the restricted system are independent on the extension so that the anyonic purification is always sufficient. 

This anyonic purification allows to define multiple copies of states of the systems with an arbitrary total charge where AEE can be defined. But in general ${\bar A}\neq B$ due to the existence of $V_{ab}^c$. 
Therefore, if the total charge $c$ is nontrivial, AEE can be asymmetric even if the state is pure, that is, 
\begin{equation}
E_A^\infty(|\psi\>)\neq E_B^\infty(|\psi\>).\label{anmix}
\end{equation}
Clearly in this case, Theorem 2 does not hold anymore. 
In particular, two independent pure states with non-abelian total charges can behave like a mixed state. 
This is due to the fact that we cannot access the reference system of the anyonic purification and the total charge of the combined state is a superposition of different charges. For this reason, the analysis of bipartite entanglement for these ``anyonically mixed'' states is more subtle.


\begin{thebibliography}{10}

\bibitem{PhysRevLett.48.1559}
D.~C. {Tsui}, H.~L. {Stormer}, and A.~C. {Gossard},
\newblock {\em Phys. Rev. Lett.}, 48:1559--1562, 1982.

\bibitem{PhysRevLett.50.1395}
R.~B. {Laughlin},
\newblock {\em Phys. Rev. Lett.}, 50:1395--1398, 1983.

\bibitem{PhysRevLett.53.722}
D.~{Arovas}, J.~R. {Schrieffer}, and F.~{Wilczek},
\newblock {\em Phys. Rev. Lett.}, 53:722--723, 1984.

\bibitem{Wenrev2013}
X.~G.~{Wen},
\newblock {\em ISRN Cond. Mat. Phys.}, 2013, 198710, 2013.

\bibitem{RevModPhys.81.865}
R.~{Horodecki}, P.~{Horodecki}, M.~{Horodecki}, and K.~{Horodecki},
\newblock {\em Rev. Mod. Phys.}, 81:865--942, 2009.

\bibitem{PhysRevLett.96.110404}
A.~{Kitaev} and J.~{Preskill},
\newblock {\em Phys. Rev. Lett.}, 96:110404, 2006.

\bibitem{PhysRevLett.96.110405}
M.~{Levin} and X.-G. {Wen},
\newblock {\em Phys. Rev. Lett.}, 96:110405, 2006.

\bibitem{2008JHEP...05..016D}
S.~{Dong}, E.~{Fradkin}, R.~G. {Leigh}, and S.~{Nowling},
\newblock {\em J. High Energy Phys.}, 5:16, 2008.

\bibitem{PhysRevLett.111.220402}
B.~J. {Brown}, S.~D. {Bartlett}, A.~C. {Doherty}, and S.~D. {Barrett},
\newblock {\em Phys. Rev. Lett.}, 111:220402, 2013.

\bibitem{2014arXiv1403.0702H}
S.~{He}, T.~{Numasawa}, T.~{Takayanagi}, and K.~{Watanabe},
\newblock {\em Phys. Rev. D}, 90:041701, 2014.

\bibitem{2008AnPhy.323.1729H}
K.~{Hikami},
\newblock {\em Ann. Phys.}, 323:1729--1769, 2008.

\bibitem{PhysRevLett.70.1895}
C.~H.~{Bennett}, G.~{Brassard}, C.~{Cr\'epeau}, R.~{Jozsa}, A.~{Peres} and W.~K.~{Wootters},
\newblock{\em Phys. Rev. Lett.}, 70:1895, 1993.

\bibitem{PhysRevLett.86.5188}
R.~{Raussendorf} and H.~J.~{Briegel},
\newblock{\em Phys. Rev. Lett.}, 86:5188, 2001.

\bibitem{Kitaev2003a}
A.~Y. {Kitaev},
\newblock {\em Ann. Phys.}, 303:2--30, 2003.

\bibitem{Freedman2002}
M.~H. {Freedman}, M.~{Larsen}, and Z.~{Wang},
\newblock {\em Comm. Math. Phys.}, 227(3):605--622, 2002.

\bibitem{RevModPhys.80.1083}
C. Nayak, S.~H. Simon, A. Stern, M. Freedman, and S. D.~Sarma,
\newblock {\em Rev. Mod. Phys.}, 80:1083--1159, 2008.

\bibitem{PhysRevLett.91.097903}
S.~D. {Bartlett} and H.~M. {Wiseman},
\newblock {\em Phys. Rev. Lett.}, 91:097903, 2003.

\bibitem{PhysRevLett.92.087904}
N.~{Schuch}, F.~{Verstraete}, and J.~I. {Cirac},
\newblock {\em Phys. Rev. Lett.}, 92:087904, 2004.

\bibitem{PhysRevA.87.022338}
N.~{Friis}, A.~R. {Lee}, and D.~E. {Bruschi},
\newblock {\em Phys. Rev. A}, 87:022338, 2013.

\bibitem{donald2002uniqueness}
M.~J. {Donald}, M.~{Horodecki}, and O.~{Rudolph},
\newblock {\em J. Math. Phys.}, 43:4252, 2002.

\bibitem{1974CS}
S.-S.~{Chern} and J.~{Simons},
\newblock {\em Ann. Math.}, 99(1):pp. 48--69, 1974.

\bibitem{Wittentqft}
E.~{Witten},
\newblock {\em Comm. Math. Phys.}, 121(3):351--399, 1989.

\bibitem{PhysRevB.89.035105}
R.~N.~C. {Pfeifer}, 
\newblock {\em Phys. Rev. B}, 89:035105, 2014.

\bibitem{Verlinde1988360}
E.~{Verlinde},
\newblock {\em Nucl. Phys. B}, 300(0):360 -- 376, 1988.

\bibitem{Preskill2004}
J.~{Preskill},
\newblock {Lacture notes on quantum computation: Topological quantum
  computation}, 2004.
\newblock \url{http://www.theory.caltech.edu/~preskill/ph219/topological.ps}.

\bibitem{kitaev2004superselection}
A.~Y. {Kitaev}, D.~{Mayers}, and J.~{Preskill},
\newblock {\em Phys. Rev. A}, 69(5):052326, 2004.

\bibitem{2008AnPhy.323.2709B}
P.~{Bonderson}, K.~{Shtengel}, and J.~K. {Slingerland},
\newblock {\em Ann. Phys.}, 323:2709--2755, 2008.

\bibitem{CoverThomas199108}
T.~M. {Cover} and J.~A. {Thomas},
\newblock {\em Elements of Information Theory 2nd Edition}.
\newblock (Wiley-Interscience, New York, 1991).

\bibitem{Wilde201304}
M.~M. {Wilde},
\newblock {\em Quantum information theory.}
\newblock (Cambridge University Press, Cambridge, England, 2013).

\bibitem{PhysRevLett.91.027901}
S.~D. {Bartlett}, T.~{Rudolph}, and R.~W. {Spekkens},
\newblock {\em Phys. Rev. Lett.}, 91:027901, 2003.


\bibitem{PhysRevA.53.2046}
C.~H. {Bennett}, H.~J. {Bernstein}, S.~{Popescu}, and B.~{Schumacher},
\newblock {Concentrating partial entanglement by local operations}.
\newblock {\em Phys. Rev. A}, 53:2046--2052, 1996.

\end{thebibliography}
\end{document}